\documentclass[showpacs,twocolumn,preprintnumbers,amsmath,amssymb]{revtex4-1}

\usepackage{graphicx}
\usepackage{dcolumn}
\usepackage{bm}
\usepackage{color}
\usepackage{amsmath}
\usepackage{upgreek}

\begin{document}
\title{Generalized spatial differentiation from spin Hall effect of light}
\author{ Tengfeng Zhu$^{1, \dag}$, Yijie Lou$^{1, \dag } $, Yihan Zhou$^{1}$, Jiahao Zhang$^{1}$, Junyi Huang$^{1}$, Yan Li$^{3}$,  Hailu Luo$^{4}$,  Shuangchun Wen$^{4}$, Shiyao Zhu$^{1,2}$, Qihuang Gong$^{3}$, Min Qiu$^{2}$, Zhichao Ruan$^{1,2}$}
\affiliation{$^1$ Department of Physics, Zhejiang University, Hangzhou 310027, China \\
$^2$ State Key Laboratory of Modern Optical Instrumentation and College of Optical Engineering, Zhejiang University, Hangzhou 310027, China \\
$^3$ State Key Laboratory for Mesoscopic Physics and School of Physics, Peking University, Beijing 100871, China \\
$^4$ Key Laboratory for Micro/Nano Optoelectronic Devices of Ministry of Education, School of Physics and Electronics, Hunan University, Changsha 410082, China
}

\begin{abstract}
Optics naturally provides us with some powerful mathematical operations. Here we experimentally demonstrate that during reflection or refraction at a single optical planar interface, the optical computing of spatial differentiation can be realized by analyzing specific orthogonal polarization states of light. We show that the spatial differentiation is intrinsically due to the spin Hall effect of light and generally accompanies light reflection and refraction at any planar interface, regardless of material composition or incident angles. The proposed spin-optical method takes advantages of a simple and common structure to enable vectorial-field computation and perform edge detection for ultra-fast and energy-efficient image processing.
\end{abstract}

\maketitle

Planar interfaces between two isotropic materials are the simplest optical structures but fundamentally reveal the nature of wave optics, such as Snell's law, Fresnel coefficients, and Brewster angle. Several decades ago it was observed that optical beams totally reflected by planar interfaces exhibit different transverse shifts dependent on polarization state \cite{Fedorov1955, Imbert1972}. However, until recently such an effect was first theoretically studied in the context of geometric phase (Berry phase) \cite{onoda2004hall,bliokh2004modified,bliokh2009goos,bliokh2004topological}, and later as total angular momentum conservation \cite{bliokh2006conservation,bliokh2013goos}. These theories greatly deepen the understanding of the spin Hall effect (SHE) of light: The spin-dependent transverse shift is geometrically protected and generally accompanies light reflection and refraction at planar interfaces. Also they boost up experimental investigations of SHE of light in a variety of different optical systems \cite{hosten2008observation,bliokh2008geometrodynamics,qin2011observation,zhou2014observation,qin2010spin,bliokh2016spin,luo2011enhancing,wang2013spin,yin2013photonic,kapitanova2014photonic,zhou2012identifying}  and the other general spin-dependent transportation phenomena \cite{bliokh2015spin,aiello2015transverse,cardano2015spin,xiao2016spin,ling2017recent}. In particular, Hosten and Kwiat identified the connection between optical SHE of photon and classical light and proposed ``weak measurement'' method for attaining ultra-high sensitivity to angstrom scale displacements \cite{hosten2008observation}. We note that currently extensive investigations of the SHE of light are focused on analyzing non-orthogonal polarization states between incident and reflected (refracted) light in order to realize weak measurement. Few studies are involved in orthogonal polarization analyzing, which creates profoundly different output beam profile from the original one \cite{duck1989sense,bliokh2016spin,zhou2014observation,qin2011observation}.

In this Letter, we experimentally demonstrate that under paraxial approximation, by analyzing specific orthogonal polarization states, the beam profiles reflected and refracted at a single optical planar interface correspond to spatial differentiation of incident field. Although the resulted spatial differentiation seems counter-intuitive to the common knowledge that no light can pass through two orthogonal polarizers, we show that the spatial differentiation is intrinsically due to the SHE of light and occurs at any planar interface, regardless of material composition or incident angles.

Moreover, the spin-optical spatial differentiation scheme offers a robust image processing to extract the boundary of objects, where two different images can be stored in two different polarization states. We note that the optical analog computing of spatial differentiation has been of great interest because it enables massively parallel processing to an entire image in a single shot. The high-throughput operation is much more energy-efficient than the standard digital electronic processing \cite{Solli2015,caulfield2010future} and attractive for real-time image processing in medical and satellite applications \cite{PhamXuPrince00,HolyerPeckinpaugh89}.

Traditionally, such an optical analog computing in spatial domain uses a bulky system of Fourier lenses and filters in linear or nonlinear optics \cite{goodman2008introduction,qiu2018spiral}. Recently, significant efforts have been taken to shrink the thickness of such computing elements down to a single-wavelength or even sub-wavelength scale \cite{Silva2014performing,Abdollah15,Chizari2016,hwang2018plasmonic,saba2018two,pors2014analog,HwangDavis16,doskolovich2014spatial,Golov15,Youssefi16,ruan2015spatial,zhu2017plasmonic,Fang2017On,guo2018photonic,bykov2018first,dong2018optical}. In particular, Silva {\it{et al.}}  theoretically proposed a complex array of meta-atoms to realize desired mathematical operations \cite{Silva2014performing}. Further within the sub-wavelength scale, it was experimentally demonstrated that a 50nm-thick silver layer can realize spatial differentiation by exciting surface plasmon polariton \cite{zhu2017plasmonic}. However due to the limitation on polarization, all of the current miniaturization proposals \cite{Silva2014performing,Abdollah15,Chizari2016,hwang2018plasmonic,saba2018two,pors2014analog,HwangDavis16,doskolovich2014spatial,Golov15,Youssefi16,ruan2015spatial,zhu2017plasmonic,Fang2017On,guo2018photonic,bykov2018first,dong2018optical} can only process scalar field and none of them have explored polarization degree of freedom yet. Here, without any Fourier lens, the spin-optical scheme offers a simple but powerful mechanism to process vectorial field with even thin and common structure, a single planar interface. Furthermore unlike the miniaturization methods based on resonant effects, such as the surface plasmonic method \cite{zhu2017plasmonic}, the SHE of light is a non-resonant effect. Correspondingly the frequency bandwidth of the spatial differentiation computing is unlimited, which enables optically fast operation speed.

To understand the generalized spatial differentiation, we first consider spin-1 photons through a preparation-postselection process. The initial state is $\left| {{\Psi _{\rm{in}}}} \right\rangle  = \left| {{\varphi_{\rm{in}}}} \right\rangle \left| s \right\rangle $, where $\left| {{\varphi _{\rm{in}}}} \right\rangle $ denotes the initial wave function $\left| {{\varphi _{\rm{in}}}} \right\rangle  = \int {d{k_y}{{\tilde \varphi }_{\rm{in}}}\left( {{k_y}} \right)} \left| {{k_y}} \right\rangle = \int {dy{\varphi _{\rm{in}}}\left( y \right)} \left| y \right\rangle $  and $\left| s \right\rangle $  corresponds to the particle's spin state. Due to the coupling between the spin and the transverse momentum, the spin Hall effect represents the wave packet split for the parallel spin state $\left|  +  \right\rangle $  (left-circularly polarized, $s=+1$)  and the antiparallel spin state  $\left|  -  \right\rangle $ (right-circularly polarized, $s=-1$ ) with shift in opposite direction: $\left| {{\Psi _{\rm{out}}}} \right\rangle  = \hat U\left| {{\Psi _{\rm{in}}}} \right\rangle  = \int {d{k_y}\tilde \varphi_{\rm{in}} \left( {{k_y}} \right)\exp \left( {i{k_y}{{\hat \sigma}_3}\delta } \right)\left| {{k_y}} \right\rangle } \left|  \pm  \right\rangle  = \int {dy\varphi_{\rm{in}} \left( {y \pm \delta } \right)\left| y \right\rangle } \left|  \pm  \right\rangle$. Here, $\hat U$  is the evolution operator, ${\hat \sigma _3}$ is the Pauli matrix along $z$ and ${\hat \sigma _3}\left|  \pm  \right\rangle  =  \pm \left|  \pm  \right\rangle $, $\exp \left( {i{k_y}{{\hat \sigma }_3}\delta } \right)$  represents a coupling between the spin and the transverse momentum of the spin particles, the wave packets of the parallel and antiparallel spin states shift $-\delta $ and $ + \delta $, respectively.

In a special case, when we prepare the initial spin state as $\left| u \right\rangle  = \frac{1}{{\sqrt 2 }}\left( {\left|  +  \right\rangle  + \left|  -  \right\rangle } \right)$ and postselect the orthogonal spin state $\left| v \right\rangle  = \frac{1}{{\sqrt 2 i}}\left( {\left|  +  \right\rangle  - \left|  -  \right\rangle } \right)$, the final measured wave function is left as
\begin{equation}
\begin{aligned}
\left| {{\varphi _{\rm{out}}}} \right\rangle  &= \left\langle v \right| {\hat U} \left| {{\varphi _{\rm{in}}}} \right\rangle \left| u \right\rangle \\
 &= \frac{i}{2}\int {d{k_y}{{\tilde \varphi }_{in}}\left( {{k_y}} \right)\left( {{e^{i{k_y}\delta }} - {e^{ - i{k_y}\delta }}} \right)\left| {{k_y}} \right\rangle } \\
 &= \frac{i}{2}\int {dy} [{\varphi _{\rm{in}}}\left( {y + \delta } \right) - {\varphi _{\rm{in}}}\left( {y - \delta } \right)]\left| y \right\rangle
\label{eq:1}
\end{aligned}
\end{equation}
Therefore, if the initial wave function profile is much larger than the shift $\delta $, the final output wave function $\left| {{\varphi _{\rm{out}}}} \right\rangle $ is approximately proportional to the first-order spatial differentiation of the input wave function:
\begin{equation}
\left| {{\varphi _{\rm{out}}}} \right\rangle   \simeq   i\delta \int {dy} \xi (y)\left| y \right\rangle
\label{eq:2}
\end{equation}
with $\xi (y) = d{\varphi _{in}}\left( y \right)/dy$. As shown in Eq.~(\ref{eq:1}), the spatial differentiation results from the opposite shifts between the parallel and antiparallel spin states, and the postselected state is orthogonal to the initial one in order to enable destructive interference. Straightforwardly, such a spatial differentiation can also be achieved by preparing the initial spin state $\left| v \right\rangle $ and postselecting the orthogonal spin state $\left| u \right\rangle $.

\begin{figure}
\centerline{\includegraphics[width=3.2in]{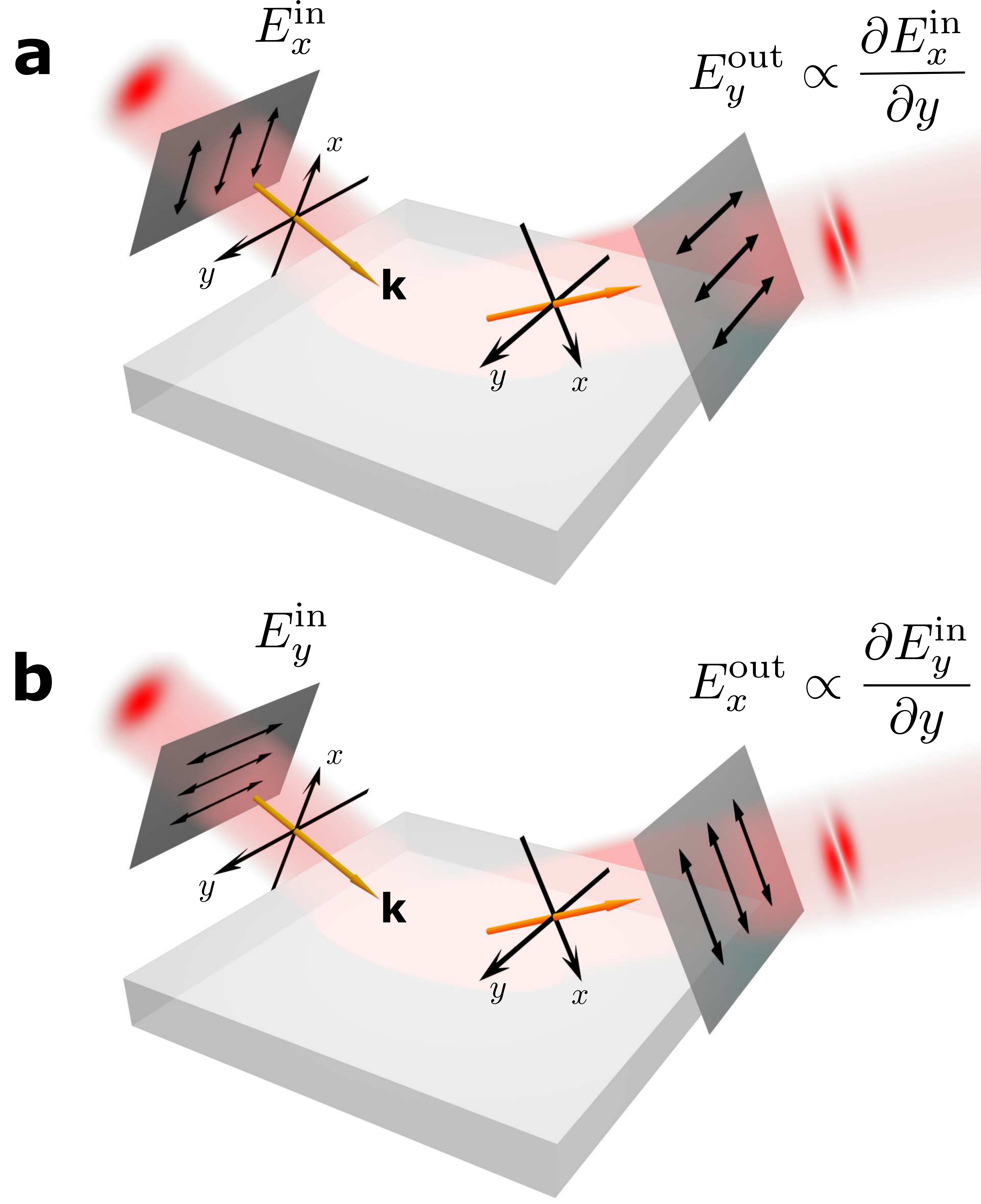}}
\caption{\label{fig:1} Schematic of spatial differentiation from the SHE of light on an optical planar interface between two isotropic materials, e.g. an air-glass interface. The two polarizers (dark grey) enable the orthogonal analyzing with the polarization indicated with double-head arrows: (a) preparing along $x$ and analyzing along $y$, (b) vice versa. As a result, when an obliquely incident paraxial beam has an electric field distribution of $f(x,y)$, the output field distribution corresponds to $\partial f/\partial y$. Here $x$ and $y$ are the beam profile coordinates for the incident and reflected light, which are perpendicular to the beam propagation direction as the wavevectors $\bf{k}$ and share the same origin on the interface, and $x$ is in the incident plane.}
\end{figure}

The principle of the spatial differentiation from the photonic SHE can be adopted in the classical level with a large number of photons in a quantum-mechanical coherent state, where each photon behaves independently and the light is treated coherently in the paraxial regime. To show the generality of the spatial differentiation, we consider a single optical interface between two isotropic media shown in Fig.~\ref{fig:1}. Since the SHE of light appears in both reflection and refraction \cite{hosten2008observation,zhou2014observation}, below our discussion focuses on the reflection case and the conclusion can be straightforwardly extended to the refraction case.

Suppose that a paraxial beam obliquely illuminates on the interface and the incident and reflected beams have the vectorial electric fields ${{\bf{E}}_{\rm{in}}}$  and ${{\bf{E}}_{\rm{out}}}$ dominating in the transversal direction. We first consider the incident and reflected electric fields along $x$ and $y$, respectively, as schematically shown in Fig.~\ref{fig:1}(a): ${{\bf{E}}_{\rm{in}}} = {{\bf{u}}_x}E_x^{\rm{in}}(x,y)$ and ${{\bf{E}}_{\rm{out}}} = {{\bf{u}}_y}E_y^{\rm{out}}(x,y)$, where ${{\bf{u}}_x}$ and ${{\bf{u}}_y}$ correspond to the $x$- and $y$- unit vectors, respectively. By vectorial spatial Fourier transform, the incident (reflected) beam can be written as the superposition of plane waves,  $E_x^{\rm{in}}$  and $E_y^{\rm{out}}$ are written by $E_{x(y)}^{\rm{in(out)}} = \iint {\tilde E_{x(y)}^{\rm{in(out)}}({k_x},{k_y})\exp (i{k_x}x)\exp (i{k_y}y)d{k_x}d{k_y}}$. Due to the continuous condition of the tangential wavevector along the interface, the incident plane wave with (${k_x}$,${k_y}$) only generates the reflection plane wave with the same (${k_x}$,${k_y}$). Therefore, the spatial transform between the incident and reflected electric fields is determined by a spatial spectral transfer function $H({k_x},{k_y}) \equiv \tilde E_y^{\rm{out}}({k_x},{k_y})/\tilde E_x^{\rm{in}}({k_x},{k_y})$. Under the paraxial approximation, the spatial spectral transfer function is simplified as (see the detail derivation in Supplemental Material (SM))
\begin{equation}
H= \frac{i \left({{r_s} +{r_p}}\right)}{{4}}({e^{ i{k_y}\delta }} - {e^{ - i{k_y}\delta }}),
\label{eq:4}
\end{equation}
where ${r_p}$ and ${r_s}$ are the Fresnel's reflection coefficients of the $p$- and $s$-polarizations for the beam incident angle ${\theta _0}$. In Eq.~(\ref{eq:4}), the two different exponential terms describe the different geometric phases experienced by the left- and the right-handed circular polarizations and resulting in the opposite transverse shifts \cite{bliokh2006conservation,bliokh2007polarization,bliokh2013goos}, and $\delta  = 2\cot {\theta _0}/{k_0}$ corresponds to the transverse shift, where ${k_0}$ is the wavevector in the incident medium and the factor of $2$ is contributed from the twice coordinate basis rotations during the reflection process. Importantly, the minus sign between two exponential terms indicates the destructive interference between the electric field components with analyzing polarization orthogonal to the initial one. Further with the requirement of $ \delta \left| {k_y} \right|  \ll  1$, Eq.~(\ref{eq:4}) is approximated as
\begin{equation}
H  \simeq  -\frac{\delta \left({{r_s}+ {r_p}}\right)}{{2}} {k_y},
\label{eq:5}
\end{equation}
which is the transfer function of a first-order $y$-directional spatial differentiator. Correspondingly, in the spatial domain, the reflected $E_y^{\rm{out}}(x,y)$ field has
\begin{equation}
E_y^{\rm{out}}= \frac{i \delta \left({r_s} + {r_p}\right)} {2} \frac{{\partial E_x^{\rm{in}}}}{{\partial y}}
\label{eq:6}
\end{equation}
Equation~(\ref{eq:6}) shows that indeed the spatial differentiation can be realized via the spin Hall effect of light, by a single optical interface. Also it is straightforward to show that when switching the orthogonal analyzing as preparing the $y$-polarized incident fields ${{\bf{E}}_{\rm{in}}} = {{\bf{u}}_y}E_y^{\rm{in}}$ and analyzing along $x$-direction ${{\bf{E}}_{\rm{out}}} = {{\bf{u}}_x}E_x^{\rm{out}}$ [Fig.~\ref{fig:1}(b)], the reflected electric field corresponds to the spatial differentiation of  the incident one as
\begin{equation}
E_x^{\rm{out}} = \frac{i \delta \left({r_p} + {r_s}\right)}{2} \frac{{\partial E_y^{\rm{in}}}}{{\partial y}}
\label{eq:7}
\end{equation}
The invariant transverse direction of the spatial differentiation after switching the orthogonal polarizations can be understood since the circular-polarization dependent shifts of the SHE occur in the transverse direction.

\begin{figure}[htbp]
\centerline{\includegraphics[width=\linewidth]{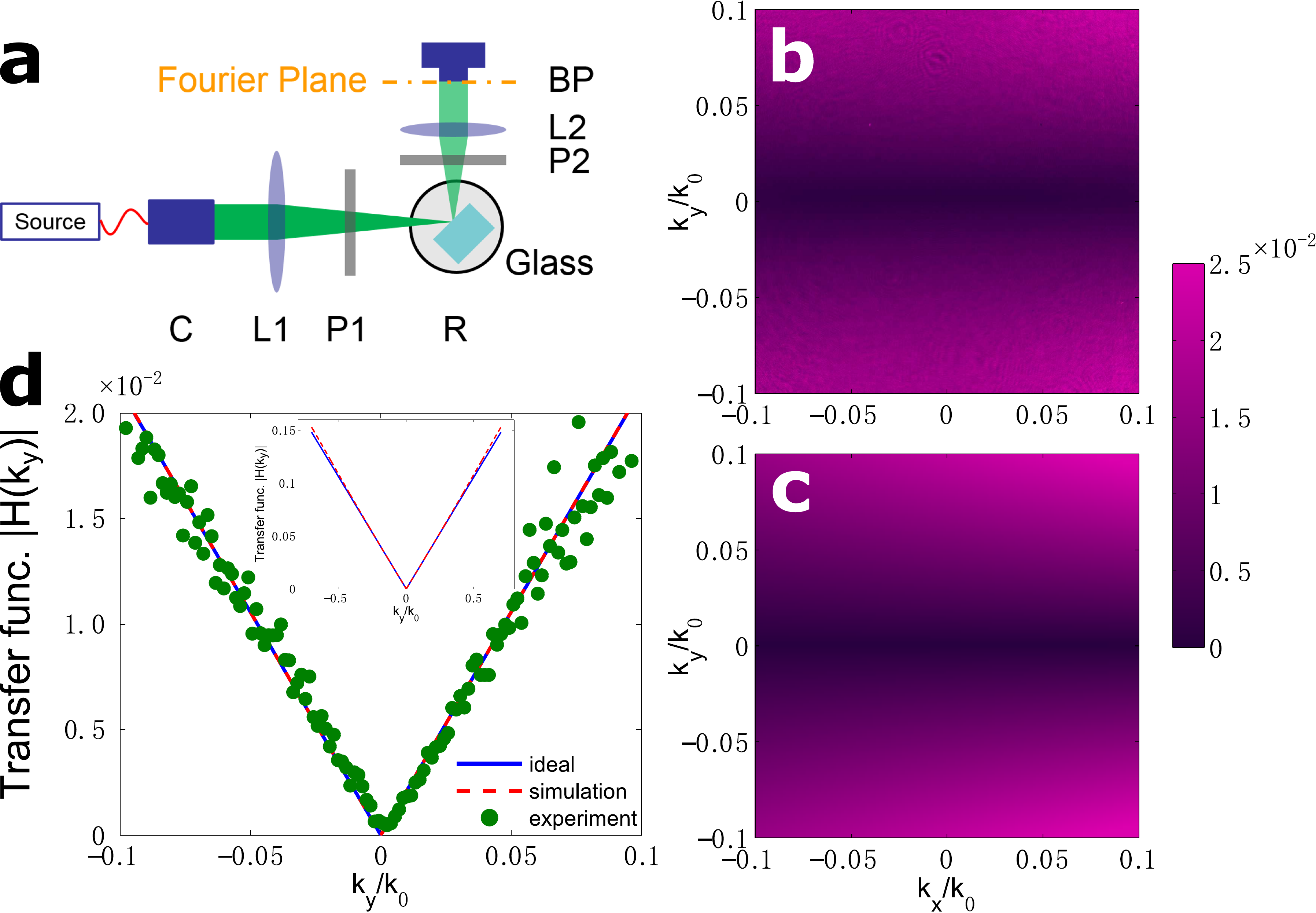}}
\caption{\label{fig:2} Measurement of the spatial spectral transfer function on an air-glass interface. (a) Experimental setup: Glass slab (material BK7); R, precision rotator; P1 and P2, polarizers; L1 and L2, lenses with focal lengths 50 and 30 mm, respectively; C, collimator; BP, beam profiler (Ophir SP620). The light source is a green laser (wavelength $\lambda_0=532$nm) and connected to the collimator through a fiber with polarization controller. (b) Measured spatial transfer function spectra. (c) Theoretical results by numerical simulations. (d) Experimental (dotted line), theoretical (dashed lines) spatial spectral transfer function for $k_x=0$, and the ideal one (solid lines) based on the first-order differentiation of Eq.~(\ref{eq:5}). The inset corresponds to a wide range from $k_y=-0.5k_0$ to $k_y=0.5k_0$. }
\end{figure}

To demonstrate the spatial differentiation, we first measure the spatial spectral transfer function on an air-glass interface, with the orthogonal analyzing. The experiment setup is schematically shown in Fig.~\ref{fig:2}(a), where an incident collimated Gaussian laser beam, with the wavelength of $532$nm, focuses on a BK7 glass surface with Lens L1, and the refractive index of the BK7 glass is 1.5195. Lens L2 projects the reflected field to the Fourier space at the back focal plane, which is measured by a beam profiler. The polarizer P1 is used to prepare the initial field polarization state, and the other one P2 is to analyze the output polarization state. The details of the method is referred to the SM. We note that two polarizers are placed between L1 and L2 in order to avoid the polarization rotation induced by the geometric phase during light focusing and collimating \cite{bliokh2011spin,bomzon2007space}.

Figure~\ref{fig:2}(b) shows the measured spatial spectral transfer function for the incident angle $\theta_0={45^ \circ }$, where the orthogonal analyzing is along $x$ and $y$-direction, respectively. The measured spatial spectral transfer function is shown in the ranges of $\left| k_{x} \right| < 0.1{k_0}$ and $\left| k_{y} \right| < 0.1{k_0}$, which are limited by the numerical aperture of the optical system. For comparison, Fig.~\ref{fig:2}(c) corresponds to the theoretical spatial transfer function by numerical simulation (see the details in SM).  Figs.~\ref{fig:2}(b) and (c) show that the experimentally measured spectral transfer function coincides with the theoretical one and has the minimum at $k_y=0$. To clearly show the effect of the spatial differentiation, the experimental and numerical transfer functions for ${k_x} = 0$ are compared with the ideal first-order differentiation based on Eq.~(\ref{eq:5}). Indeed, the spatial transfer functions agree well and exhibit a linear dependence of $k_y$ around $k_y=0$ [Fig.~\ref{fig:2}(d)]. More importantly, the inset shows that the spatial spectrum bandwidth of the $y$-direction differentiation is as large as about $\left| k_{y} \right| < 0.5k_0$.

\begin{figure}[htb!]
\centerline{\includegraphics[width=\linewidth]{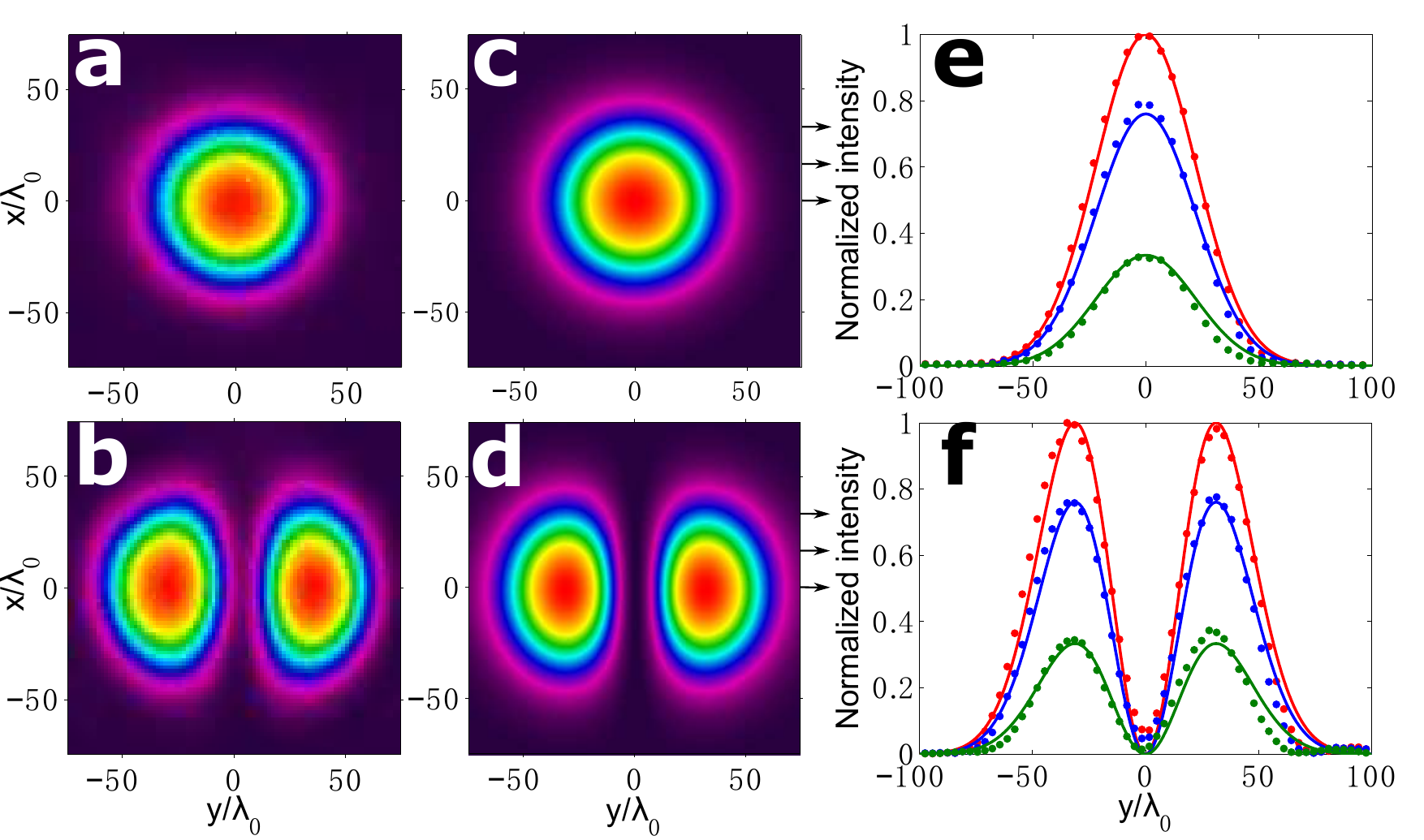}}
\caption{\label{fig:3} Spatial differentiation demonstration for a Gaussian illumination. (a,b) Measured intensity profiles of the incident and reflected beams, respectively. (c,d) Numerical Gaussian fitting (c) to the incident beam with a waist radius ${w_0} = 44.2{\lambda _0}$, and the ideal spatial differentiation results (d) corresponding to $y$-direction differentiation of (c). (e,f) Normalized intensities of the incident and reflected fields at $x=0.0{\lambda _0}$, $16.54{\lambda _0}$, and $33.08{\lambda _0}$, as indicated by the right arrows in (c) and (d), where the dotted and solid lines correspond to the intensities of the experimental and ideal results, respectively.}
\end{figure}

To illustrate the spin-optical spatial differentiation effect, we measure the reflected field distribution under a Gaussian beam illumination. Figs.~\ref{fig:3}(a) and (b) show the measured intensity profiles for the incident and reflected beams, respectively. Indeed the reflected beam exhibits a first-order Hermite-Gaussian profile with a minimum amplitude at $y=0$, which corresponds to the $y$-direction spatial differentiation of the incident Gaussian field. To quantitatively illustrate the performance of spatial differentiation, we numerically fit the incident beam with a Gaussian profile of a waist radius ${w_0} = 44.20{\lambda _0}$ as illustrated in Fig.~\ref{fig:3}(c). Fig.~\ref{fig:3} (d) shows the ideal first-order differentiation of Fig.~\ref{fig:3} (c) in the $y$-direction. In Fig.~\ref{fig:3}(e) and (f), we compare the normalized experimental results with the ideal ones at three different cut-through positions. The dotted and solid lines correspond to the intensities of the experimental and ideal results, respectively. As a direct demonstration of spatial differentiation, the experimental reflected fields show a good agreement with the ideal spatial differentiation in both the peak positions and values [Fig.~\ref{fig:3}(f)].

\begin{figure}
\centerline{\includegraphics[width=3.2in]{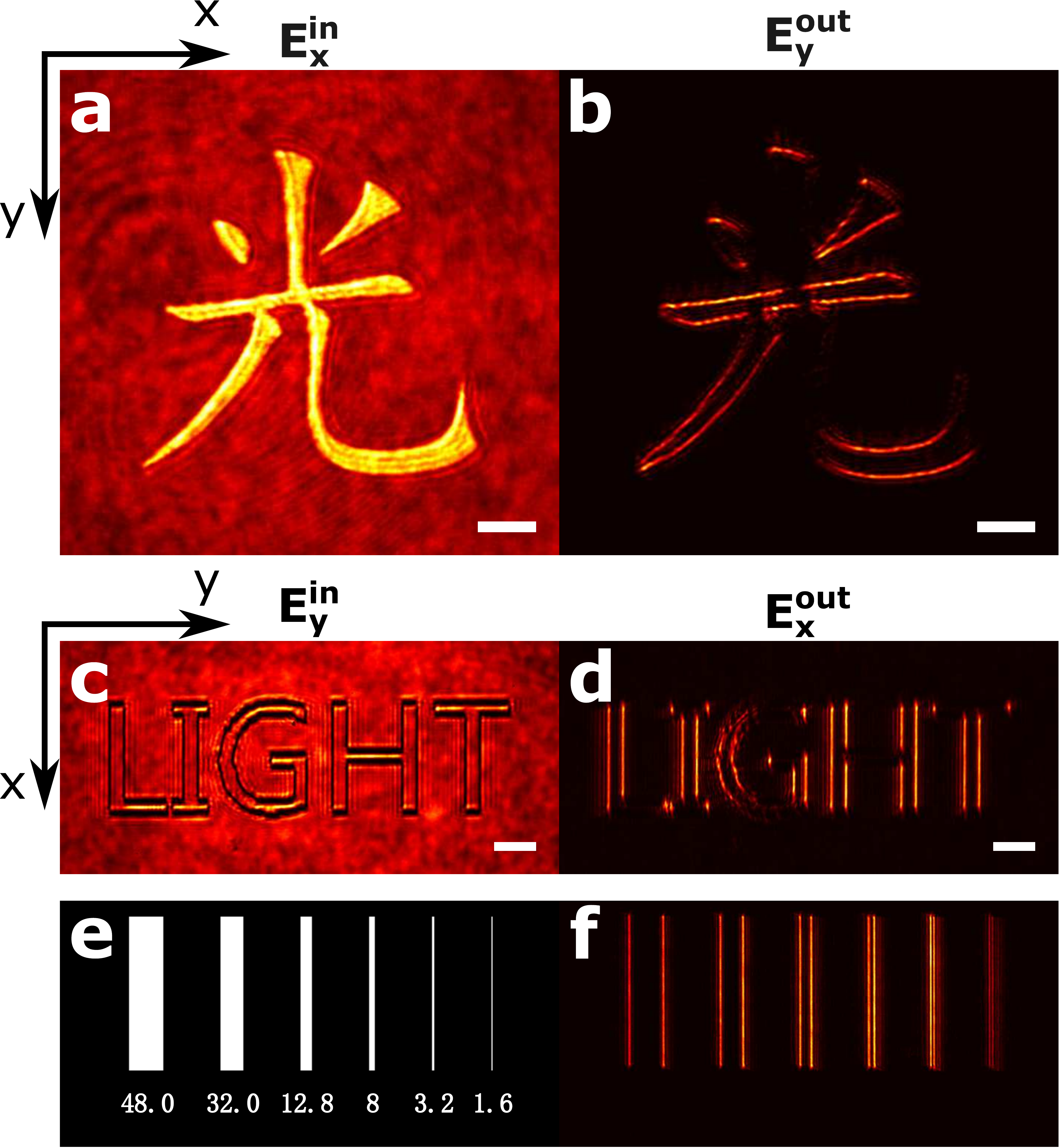}}
\caption{\label{fig:5} Edge detection for different images stored in $E^{\rm{in}}_x$ and $E^{\rm{in}}_y$, respectively, with either amplitude or phase modulation. (a) Incident image consisting of a Chinese character of light with amplitude modulation on $E^{\rm{in}}_x$. (b) Reflected intensity image corresponding to (a) by measuring $E^{\rm{out}}_y$. (c) Incident image consisting of the LIGHT letters generated with phase modulation on $E^{\rm{in}}_y$, where the inside and the outside of the letters have different phases but the same intensity. (d) Reflected intensity image corresponding to (c) by measuring $E^{\rm{out}}_x$. The white bars correspond to the length of $50$ ${\rm \upmu m}$. (e) Slot test patterns on the SLM with the different phases for the black and the white areas. The widths of the slots list below in ${\rm \upmu m}$. (f) Measured reflected intensity image corresponding to (e). }
\end{figure}	

Since the optical computing of spatial differentiation is important for the ultra-fast image processing of edge detection, we demonstrate the spin-optical spatial differentiation in these aspects. We note that spin-optical method takes advantages of image processing on vectorial fields, which enables one more degree of freedom than all the current scalar-field schemes\cite{Silva2014performing,Abdollah15,Chizari2016,hwang2018plasmonic,saba2018two,pors2014analog,HwangDavis16,doskolovich2014spatial,Golov15,Youssefi16,ruan2015spatial,zhu2017plasmonic,Fang2017On,guo2018photonic,bykov2018first,dong2018optical}.
 As the differentiation operates on the electric field rather than the intensity, the device can be used to detect an edge either in the phase or the amplitude distribution of the incident field. To show such an effect, we use a spatial light modulator (SLM) to generate incident fields with amplitude and phase modulations, respectively (see the detailed method in SM).

Figure~\ref{fig:5}(a) shows the incident image field consisting of a Chinese character of light generated with amplitude modulation on the field component of $E^{\rm{in}}_x$, where the inside and the outside of the character have different intensities. Figure~\ref{fig:5}(b) shows the measured reflected intensity of $E^{\rm{out}}_y$. It clearly exhibits the outlines of the character with spatial differentiation. Since the differentiation is along the $y$-direction, the edges perpendicular to the $y$-direction are most visible. Furthermore as long as the edge is not completely along the $y$-direction, it can be detected in the reflected beam. Figure~\ref{fig:5}(c) shows an incident image field of LIGHT letters generated with phase modulation on the field component of $E^{\rm{in}}_y$. We note that here we have rotated the incident image to demonstrate that we can detect the vertical edges. Again, the reflected light clearly exhibits only the edges of the letters in the vertical direction by measuring $E^{\rm{out}}_x$ (Fig.~\ref{fig:5}(d)), which exactly corresponds to spatial differentiation along $y$-direction. The differentiator performs spatial differentiation along only a single direction. This directional selectivity feature is very useful in image processing to determine and classify edge directions \cite{JainKasturiSchunck95}.

We also  verify the generality of spin-optical spatial differentiation by varing different incident angles (Fig.~S3(a-f) of SM). Moreover, since the SHE of light on the interface is a geometric effect which is independent of material, we demonstrate the spatial differentiation in the total reflection case on 210nm-thick gold layer (Fig.~S3(g-l) of SM)).

As shown in Fig.~\ref{fig:2}(d), since there is a finite spatial bandwidth for the spatial differentiation, it will not be able to resolve two edges that are very close to each other. In order to show the edge detection resolution we generate slot test patterns on the SLM [Fig.~\ref{fig:5}(e)], and the corresponding reflected intensity is measured and shown in Fig.~\ref{fig:5}(f). It shows that the minimum separation between the two edges that can be resolved, is about $3.2$ $ {\rm \upmu m}$. The resolution is mainly limited by the small numerical aperture of the image system and can be further improved.

In summary, we experimentally demonstrate the generality of spatial differentiation based on the SHE of light. The observation of the spin-optical spatial differentiation is dependent on three key elements: oblique incidence, coherent paraxial beam, and polarizer. The SHE of light vanishes at normal incidence for symmetry reasons, therefore the spatial differentiation signal is decreased with a small value of $\left| {{r_s} + {r_p}} \right|$ when reducing the incident angle [c.f. Eq.~(\ref{eq:5})]. The spin-optical method exhibits high signal noise ratio of spatial differentiation, which is contributed by the orthogonal analyzing strongly reducing background noise. Moreover, we show that such a spatial differentiation takes advantages of a simple and common structure to enable vectorial-field computation. Due to the generality of SHE, the proposed method points a way to ultra-fast info processing in a variety of optical systems  and can even forward to the processing of electron beams.

%

\noindent $^\dag$These authors contributed equally to this work.

%

\clearpage

\makeatletter
\renewcommand{\thefigure}{S\arabic{figure}}
\def\tagform@#1{\maketag@@@{(\ignorespaces\textbf{S#1}\unskip\@@italiccorr)}}
\renewcommand{\eqref}[1]{\textup{{\normalfont S\ref{#1}}\normalfont}}
\makeatother

\setcounter{equation}{0}
\setcounter{figure}{0}
\section{Supplemental Materials}
\subsection{Calculation of spatial spectral transfer function for the spin-optical differentiation}

\begin{figure}
\centerline{\includegraphics[width=3.2in]{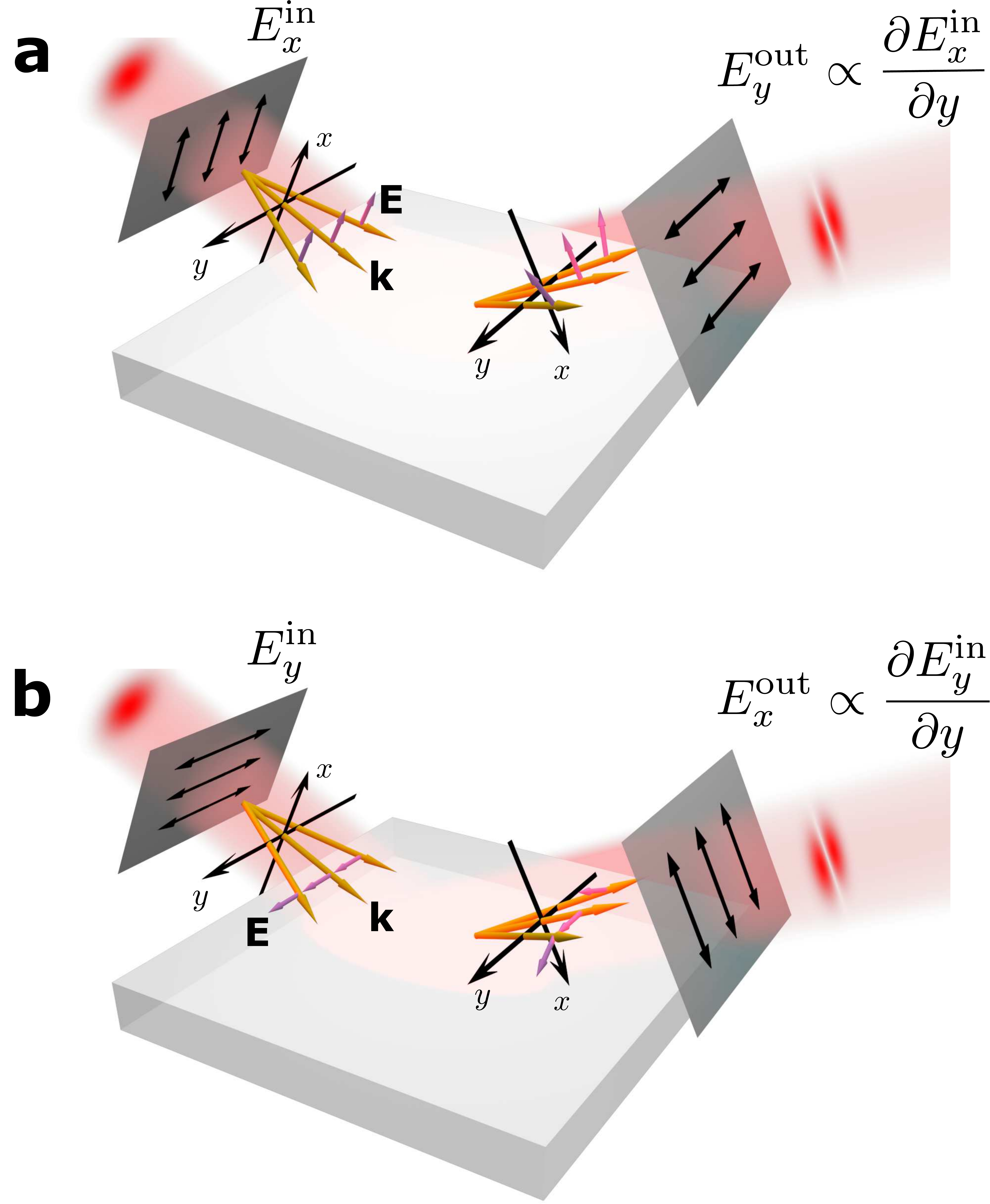}}
\caption{\label{fig:S1} Supplementary diagram for the schematic of spatial differentiation based on the SHE of light (Fig.~1 in main text). Here the different rotation angles of the electric field vectors $\bf{E}$ (purple arrows) indicate the different experienced geometric phases of the left- and right-handed circularly polarized plane waves for different wavevectors $\bf{k}$ (golden arrows) during the light reflection. The spatial differentiation results from the opposite shifts of the incident beam, and the analyzing polarization is orthogonal to the initial one in order to enable destructive interference between the shifted beams.}
\end{figure}

\begin{figure}
\centering
\includegraphics[width=\linewidth]{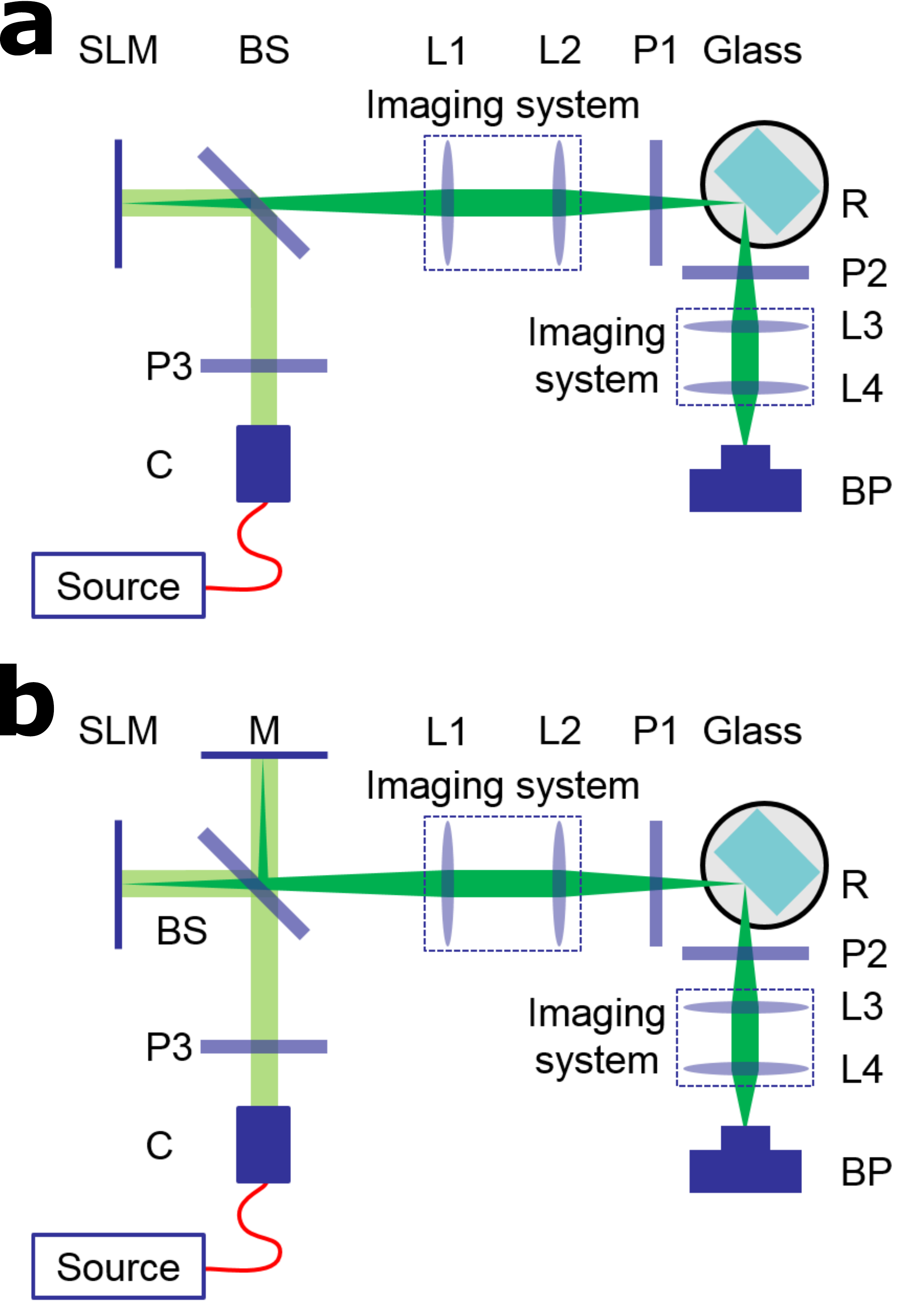}
\caption{\label{fig:setup} (a,b) Schematic of the optical system for phase and amplitude modulated field images generated with SLM, respectively. Components include C: collimator, P: polarizer, BS: beam splitter, L: lens, SLM: spatial light modulator, M: mirror, Glass slab (material BK7), R: precision rotator, BP: beam profiler (Ophir SP620). The light source is a green laser (wavelength $\lambda_0=532$nm) and connected to the collimator through a fiber with a polarization controller.}
\end{figure}

\begin{figure*}
\centerline{\includegraphics[width=6.2in]{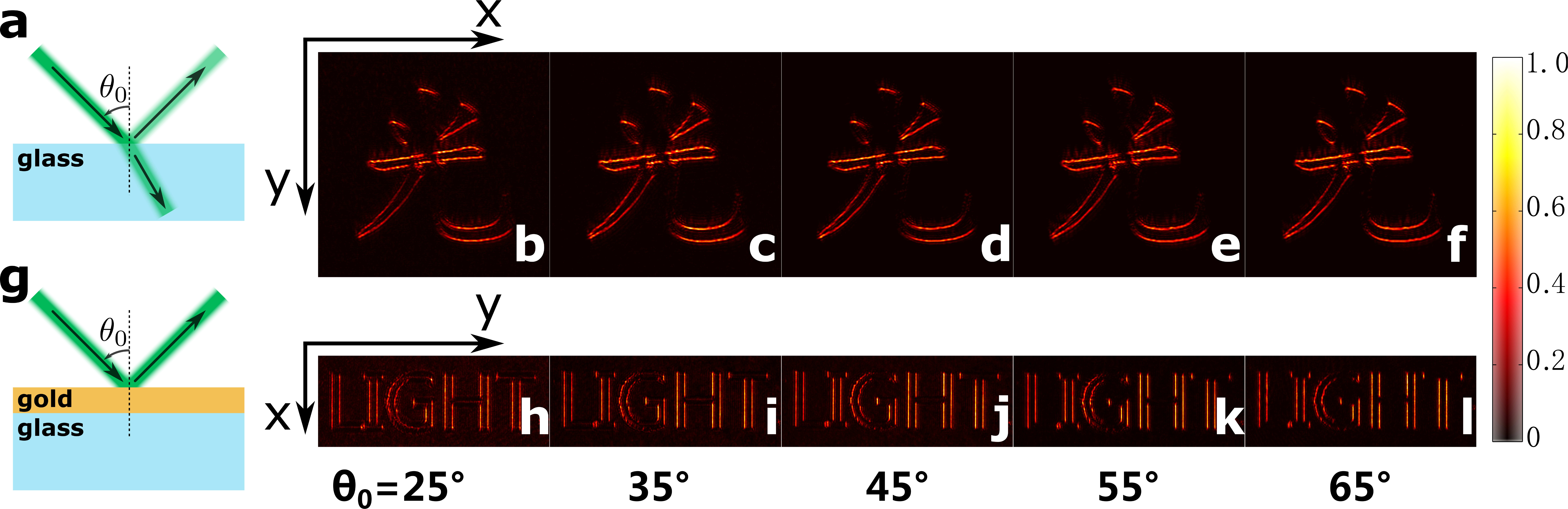}}
\caption{\label{fig:6} Edge detection demonstration on different material interfaces and for a variety of incident angles. (a-f) the weak reflection case on the air-glass interface, (g-l) the total reflection case with a planar interface on 210nm-thick gold layer. (a) and (g) are the schematics of the light reflection for two cases, respectively.  The spatial differentiation results are shown for the incident images with (b-f) the Chinese character of light and (h-l) the letters of LIGHT. Here the field intensities are normalized with each maximum. The incident angle for each column lists below. }
\end{figure*}

Considering a plane wave with a wavevector ${\bf{k}}$ as shown in Fig.~\ref{fig:S1}, we define the  $s$- and $p$-polarization as the electric field vectors along ${{\bf{u}}_{\bf{s}}} = {{{{\bf{u}}_{\bf{k}}} \times {\bf{n}}} \mathord{\left/{\vphantom {{{{\bf{u}}_{\bf{k}}} \times {\bf{n}}} {\left| {{{\bf{u}}_{\bf{k}}} \times {\bf{n}}} \right|}}} \right.\kern-\nulldelimiterspace} {\left| {{{\bf{u}}_{\bf{k}}} \times {\bf{n}}} \right|}}$ and ${{\bf{u}}_{\bf{p}}}{\bf{ = }}{{\bf{u}}_{\bf{s}}} \times {{\bf{u}}_{\bf{k}}}$, respectively. Here ${{\bf{u}}_{\bf{k}}}$  is the normalized wavevector as ${{\bf{u}}_{\bf{k}}}{\rm{ = }}{{\bf{k}} \mathord{\left/
 {\vphantom {{\bf{k}} {\left| {\bf{k}} \right|}}} \right.
 \kern-\nulldelimiterspace} {\left| {\bf{k}} \right|}}$, and ${\bf{n}}$ is the unit normal vector of the interface. Accordingly, ${{\bf{u}}_{{ + }}}$ and ${{\bf{u}}_{-}}$  are the left and right circular polarization bases for the plane wave, respectively
\begin{equation}
{{\bf{u}}_{ { \pm }}} = \frac{1}{{\sqrt 2 }}\left( {{{\bf{u}}_{\bf{p}}} \pm {i}{{\bf{u}}_{\bf{s}}}} \right).
\end{equation}
Under the paraxial approximation, the incident and reflected beams have the vectorial electric fields dominating in the transversal directions:
\begin{subequations}
\begin{equation}
\begin{aligned}
{{\bf{E}}_{{\rm{in}}}} &= {\bf{u}}_ + ^{{\bf{i0}}}E_ + ^{{\rm{in}}}\left( {x,y} \right) + {\bf{u}}_ - ^{{\bf{i0}}}E_ - ^{{\rm{in}}}\left( {x,y} \right)\\
 &= \iint {[{\bf{u}}_ + ^{{\bf{i0}}}\tilde E_ + ^{{\rm{in}}}\left( {{k_x},{k_y}} \right) + {\bf{u}}_ - ^{{\bf{i0}}}\tilde E_ - ^{{\rm{in}}}\left( {{k_x},{k_y}} \right)]{e^{i{k_x}x}}{e^{i{k_y}y}}d{k_x}d{k_y}}
\end{aligned}
\end{equation}
\begin{equation}
\begin{aligned}
{{\bf{E}}_{{\rm{out}}}} &= {\bf{u}}_ + ^{{\bf{r0}}}E_ + ^{{\rm{out}}}\left( {x,y} \right) + {\bf{u}}_ - ^{{\bf{r0}}}E_ - ^{{\rm{out}}}\left( {x,y} \right)\\
 &= \iint {[{\bf{u}}_ + ^{{\bf{r0}}}\tilde E_ + ^{{\rm{out}}}\left( {{k_x},{k_y}} \right) + {\bf{u}}_ - ^{{\bf{r0}}}\tilde E_ - ^{{\rm{out}}}\left( {{k_x},{k_y}} \right)]{e^{i{k_x}x}}{e^{i{k_y}y}}d{k_x}d{k_y}}
\end{aligned}
\end{equation}
\end{subequations}
where $x$ and $y$ are the beam coordinates as shown in Fig.~\ref{fig:S1}, and ${\bf{u}}^{{\bf{i0}}}_{\bf{ \pm }}$ and ${\bf{u}}^{{\bf{r0}}}_{\bf{ \pm }}$ correspond to the circular polarization bases for the central wavevector of the incident and reflected beams, respectively. $E_ \pm ^{\rm{in(out)}}\left( {x,y} \right)$ are the amplitude distributions in the spatial domain for incident (reflected) beam, and $\tilde E_ \pm ^{\rm{in(out)}}$ are the Fourier spectrum of the amplitudes. To calculate the spatial spectral transfer function, we decompose the vectorial fields ${{\bf{E}}_{{\rm{in}}}}$  and ${{\bf{E}}_{\rm{out}}}$ into plane waves. Futhermore, due to the continuous condition of the tangential wavevector along the interface, the incident plane wave with the transversal component $\left( {{k_x},{k_y}} \right)$ only generates the reflection plane wave with the same $\left( {{k_x},{k_y}} \right)$. Therefore, the Fourier spectrum of the reflected beam can be obtained by a matrix ${\bf{R}}$ as
\begin{equation}
\left( {\begin{array}{*{20}{c}}
{\tilde E_ + ^{\rm{out}}}\\
{\tilde E_ - ^{\rm{out}}}
\end{array}} \right) = {\bf{R}}\left( {\begin{array}{*{20}{c}}
{\tilde E_{\rm{ + }}^{\rm{in}}}\\
{\tilde E_ - ^{\rm{in}}}
\end{array}} \right) = {\bf{U}}_{\bf{2}}^\dag {\bf{\tilde R}}{{\bf{U}}_{\bf{1}}}\left( {\begin{array}{*{20}{c}}
{\tilde E_{\rm{ + }}^{\rm{in}}}\\
{\tilde E_ - ^{\rm{in}}}
\end{array}} \right)
\end{equation}
where ${{\bf{U}}_{ {1}}}$  and ${{\bf{U}}_{ {2}}}$  are the transfer matrices between the circular polarization base for the central wavevector and that for each other wavevector
\begin{subequations}
\begin{equation}
\renewcommand*{\arraystretch}{1.5}
\setlength{\arraycolsep}{10pt}
{{\bf{U}}_1} = \left( {\begin{array}{*{20}{c}}
{{{\left( {{\bf{u}}^{\bf{i}}_{\bf{ + }}} \right)}^*} \cdot {\bf{u}}^{{\bf{i0}}}_{\bf{ + }}{\rm{     }}}&{{{\left( {{\bf{u}}^{\bf{i}}_{\bf{ + }}} \right)}^*} \cdot {\bf{u}}^{{\bf{i0}}}_ - }\\
{{{\left( {{\bf{u}}^{\bf{i}}_ - } \right)}^*} \cdot {\bf{u}}^{{\bf{i0}}}_{\bf{ + }}{\rm{     }}}&{{{\left( {{\bf{u}}^{\bf{i}}_ - } \right)}^*} \cdot {\bf{u}}^{{\bf{i0}}}_ - }
\end{array}} \right)
\end{equation}
\begin{equation}
\renewcommand*{\arraystretch}{1.5}
\setlength{\arraycolsep}{10pt}
{{\bf{U}}_2} = \left( {\begin{array}{*{20}{c}}
{{{\left( {{\bf{u}}^{\bf{r}}_{\bf{ + }}} \right)}^*} \cdot {\bf{u}}^{{\bf{r0}}}_{\bf{ + }}}&{{{\left( {{\bf{u}}^{\bf{r}}_{\bf{ + }}} \right)}^*} \cdot {\bf{u}}^{{\bf{r0}}}_ - }\\
{{{\left( {{\bf{u}}^{\bf{r}}_ - } \right)}^*} \cdot {\bf{u}}^{{\bf{r0}}}_{\bf{ + }}}&{{{\left( {{\bf{u}}^{\bf{r}}_ - } \right)}^*} \cdot {\bf{u}}^{{\bf{r0}}}_ - }
\end{array}} \right)
\end{equation}
\end{subequations}
where the asterisk $*$ represents the complex conjugate operation, and ${\bf{u}}^{{\bf{i}}\left( {\bf{r}} \right)}_{\bf{ \pm }}$ denotes the incident (reflected) circular bases for the wavevector ${\bf{k}}$ with $\left( {{k_x},{k_y}} \right)$. The matrix ${\bf{\tilde R}}$ describes the reflection coefficients for the left- and right-handed  circularly polarized plane waves:
\begin{equation}
\renewcommand*{\arraystretch}{1.5}
\setlength{\arraycolsep}{10pt}
{\bf{\tilde R}}{\rm{ = }}\frac{1}{2}\left( {\begin{array}{*{20}{c}}
{{r_{ {p}}} + {r_{ {s}}}}&{{r_{ {p}}} - {r_{ {s}}}}\\
{{r_{ {p}}} - {r_{ {s}}}}&{{r_{ {p}}} + {r_{ {s}}}}
\end{array}} \right)
\end{equation}
where ${r_{{p}}}$ and ${r_{{s}}}$ are the Fresnel's reflection coefficients for each $p$- and $s$-polarized plane wave with the wavevector ${\bf{k}}$, respectively.

In order to realize the spatial differentiation, we prepare the incident field polarization state and analyze the reflected one along $x$ and $y$, respectively, as schematically shown in Fig.~\ref{fig:S1}(a).  Under the paraxial approximation, the incident and reflected electric field distributions are ${{\bf{E}}_{{\rm{in}}}} = {{\bf{u}}_{{x}}}E_x^{\rm{in}}\left( {x,y} \right)$  and ${{\bf{E}}_{\rm{out}}} = {{\bf{u}}_{{y}}}E_y^{\rm{out}}\left( {x,y} \right)$. The spatial spectral transfer function $H$ can be calculated as:
\begin{equation}
H = V_y^\dag {\bf{R}}{V_x} \label{eq:trans}
\end{equation}
where ${V_x} = \frac{1}{{\sqrt 2 }}{(\begin{array}{*{20}{c}}
1&1
\end{array})^T}$ and ${V_y} = \frac{i}{{\sqrt 2 }}{(\begin{array}{*{20}{c}}
{ - 1}&1
\end{array})^T}$ correspond to the prepared polarization along ${{\bf{u}}_{ {x}}}$  and analyzed polarization along ${{\bf{u}}_{ {y}}}$, respectively, due to ${{\bf{u}}_{ {x}}} = \frac{1}{{\sqrt 2 }}\left( {{\bf{u}}^{{\bf{i0}}}_{\bf{ + }} + {\bf{u}}^{{\bf{i0}}}_- } \right)$ and ${{\bf{u}}_{ {y}}} = \frac{i}{{\sqrt 2 }}\left( { - {\bf{u}}^{{\bf{r0}}}_{\bf{ + }} + {\bf{u}}^{{\bf{r0}}}_- } \right)$.

\subsection{Spatial spectral transfer function under the paraxial approximation}

Here we show the spatial spectral transfer function [Eq.(3) in main text] for the spin-optical differentiation. The derivation is based on different experienced geometric phases of the left- and right-handed circular plane waves, which lead to the spin-dependent transverse shift \cite{bliokh2013goos}.

Considering the paraxial approximation where the incident beam spectrum is limited as $\left| {{k_x}} \right| \ll {k_0}$  and $\left| {{k_y}} \right| \ll {k_0}$, the matrices ${{\bf{U}}_{ {1}}}$ and ${{\bf{U}}_{ {2}}}$ can be approximately evaluated through the geometric phases of the left- and right-handed circularly polarized waves
\begin{equation}
{{\bf{U}}_{\rm{j}}} = \left( {\begin{array}{*{20}{c}}
{\exp \left( {i\Phi _{\rm{B}}^{\rm{j}}} \right)}&0\\
0&{\exp \left( { - i\Phi _{\rm{B}}^{\rm{j}}} \right)}
\end{array}} \right)  \label{eq:1}
\end{equation}
where ${\rm{j}}=1,2$, $\Phi _{\rm{B}}^1 = \frac{{{k_y}}}{{{k_0}}}\cot {\theta _0}$ and $\Phi _{\rm{B}}^2 =  - \frac{{{k_y}}}{{{k_0}}}\cot {\theta _0}$ correspond to the geometric phases during the wavevector rotation \cite{bliokh2013goos}. Eq.~(S\ref{eq:1}) shows that the left- and right-handed circularly polarized light with a wavevector $\bf{k}$ experience the opposite geometric phases during the light reflection, resulting in the SHE of light. Furthermore, the Fresnel's reflection coefficients ${r_{{p}}}$ and ${r_{{s}}}$ in the matrix ${\bf{\tilde R}}$ are approximately equal to those coefficients with an incident angle ${\theta _0}$, respectively. Therefore, the matrix ${\bf{R}}$ can be written as
\begin{equation}
{\bf{R}} = \frac{1}{2}\left( {\begin{array}{*{20}{c}}
{\left( {{r_{ {p}}} + {r_{ {s}}}} \right){e^{i{k_y}\delta }}}&{{r_{ {p}}} - {r_{ {s}}}}\\
{{r_{ {p}}} - {r_{ {s}}}}&{\left( {{r_{ {p}}} + {r_{ {s}}}} \right){e^{ - i{k_y}\delta }}}
\end{array}} \right) \label{eq:2}
\end{equation}
where $\delta {\rm{ = }}\frac{2}{{{k_0}}}\cot {\theta _0}$ corresponds to the transverse shift.  Eq.~(S\ref{eq:2}) indicates that during the reflection the electric field vectors $\bf{E}$ have different rotation angles for different wavevector $\bf{k}$ [Fig.~\ref{fig:S1}]. According to Eq.~(S\ref{eq:trans}), the spatial spectral transfer function $H$ is
\begin{equation}
H = \frac{i}{4}\left( {{r_{ {p}}} + {r_{ {s}}}} \right)\left( {{e^{i{k_y}\delta }} - {e^{ - i{k_y}\delta }}} \right) \label{eq:3}
\end{equation}
Therefore, it shows that the spatial differentiation results from the opposite shifts of the incident light, and the analyzing polarization is orthogonal to the initial one in order to enable destructive interference between the shifted beams, described as the minus sign in Eq.~(S\ref{eq:3}).

\subsection{Experimental measurement of spatial spectral transfer function}

In order to perform the orthogonal analyzing [Fig.2(a) in main text], P1 was placed to make the incident field polarization state along $x$-direction. The other polarizer P2 was first along $y$-direction and the reflected spatial spectra were measured by the beam profiler at the focal plane of L2. The incident spatial spectra are obtained by removing the glass and rotating the reflection path L2 and P2 to the normal incidence. With the polarizer P2 along the same polarization as P1, we measure the incident spatial spectra at the back focal plane of L2. Correspondingly, the spatial spectral transfer function is obtained by normalizing the reflected spectrum data with the incident ones. We measure the incident and the reflected spectra with the beam profiler (Ophir SP620) instead of conventional charge-coupled devices (CCDs) because the CCD might use gamma correction and cannot accurately measure the field intensity.

\subsection{Phase/Amplitude modulated field image generated by SLM}

We generate the incident field images with a spatial light modulator (SLM: Holoeye PLUTO-NIR-011), which is a reflective-phase-only modulator. For the phase modulation, the experimental setup is schematically shown in Fig.~\ref{fig:setup}(a), where a collimated green (532nm) laser beam illuminates on the SLM after a polarizer P3. The phase-modulated image field is projected to the air-glass interface with a conjugate imaging system. In order to generate the amplitude modulated incident field with the SLM, a Michelson configuration is used to make the phase-modulated field interfere with a reference collimated wave [Fig.~\ref{fig:setup}(b)]. The incident angle is controlled with a precision rotator R. Two polarizers P1 and P2 are used to enable the spatial differentiation through preparing and analyzing the polarization states in orthogonal directions.

\subsection{Spin-optical spatial differentiation for different incident angles and materials }

To verify the generality of spin-optical spatial differentiation, we also experimentally demonstrate it for different incident angles and materials. Figures~\ref{fig:6}(b-f) show the spatial differentiation results on the air-glass interface for the image with the Chinese character of light, when the incident angle is varied from $25^\circ$ to $65^\circ$ with a step of $10^\circ$. They clearly demonstrate that the spatial differentiation occurs for all the different incident angles. Moreover, since the SHE of light on the interface is a geometric effect which is independent of material, we also demonstrate the spatial differentiation in the total reflection case on 210nm-thick gold layer. Figures~\ref{fig:6}(h-l) show the spatial differentiation results for the image consisting of the LIGHT letters with different incident angles. Indeed, these results confirm that the spatial differentiation is regardless of the composition materials of the planar interface.



%

\end{document}